\documentclass[12pt]{article}
\usepackage{amsfonts}
\usepackage{graphicx}
\usepackage[psamsfonts]{amssymb}
\usepackage[mathscr]{eucal}
\usepackage[numbers,sort&compress]{natbib}
\begin{document}
\begin{center}
{\bf Statistical field theory for a multicomponent fluid: The collective variables approach\footnote{This paper is dedicated to Professor Ivan Vakarchuk on the occasion of his 60th birthday}}
\\
[.5cm] Oksana Patsahan, Ihor Mryglod\\
[.5cm]
{\it Institute for Condensed Matter Physics, National Academy of
Sciences of Ukraine, 79011, Lviv, Ukraine}\\
[.5cm]
Jean-Michel Caillol\\
[.5cm]
{\it Laboratoire de Physique Th\'eorique
CNRS UMR 8627, B\^at. 210
Universit\'e de Paris-Sud 91405 Orsay Cedex, France}
\end{center}

\begin{abstract}
Using the collective variables (CV) method the basic relations of
statistical field theory of a multicomponent non-homogeneous fluids are reconsidered. The
corresponding CV action depends on two sets of scalar fields - fields
$\rho_{\alpha}$ connected to the local density fluctuations of the $\alpha$th species of particles and fields
$\omega_{\alpha}$ conjugated to $\rho_{\alpha}$. The explicit expressions for the CV field correlations and  their relation to the density correlation functions are found. The perturbation theory is formulated and a mean field level (MF) of the theory is considered in detail.
\end{abstract}
{\bf Pacs:} 05.70.Fh, 02.70.Rr

\newpage
\section{Introduction}
In recent years much attention has been focused on an issue of the phase transitions in multicomponent fluid mixtures, especially in  ionic fluids. In spite of significant progress in this field, such systems are far from being completely  understood. The investigation of  complex models is of great importance in understanding the nature of critical and phase behavior of real ionic fluids which demonstrate both the charge and size asymmetry. The  powerful tools for the study of multicomponent continuous systems are those based on the functional methods. In many cases the partition function of multicomponent models (see, e.g.\cite{Vakarchuk}) can be re-expressed as a functional integral after performing the Hubbard-Stratonovich transformation
\cite{stratonovich,hubbard}, a simple device proposed in the 50ies. Nearly at the same time another method, the method of collective variables (CVs), that allows in a explicit way to construct a functional
representation for many-particle interacting systems was developed
\cite{zubar,jukh}. The method, proposed initially in the 1950s \cite{bohm,zubar,jukh} for the description of the classical charged many particle systems and developed later for the needs of the phase transition theory
\cite{yuk,yuk2,patyuk3,patyuk4}, was in fact one of the first successful attempts to attack the problems of statistical physics using the functional integral representation. 
Recently, the rigorous scalar field KSSHE
(Kac-Siegert-Stratonovich-Hubbard-Edwards) theory \cite{Cai-Mol,Cai-JSP}, which uses the Stratonovich-Hubbard transformation, was developed to describe the phase equilibria in  simple and ionic fluids.
As was shown \cite{caillol_patsahan_mryglod, patsahan_mryglod_CM}, both groups of theories  are in close relation.

In \cite{caillol_patsahan_mryglod} the CV representation of simple (one-component) fluids was reexamined from the point of view of statistical field theory. Our goal here is to derive the exact functional representation for the grand canonical partition function of a non-homogeneous multicomponent fluid. We reformulate the method of CV in real space and derive the CV action that depends on two sets of scalar fields - fields $\{\rho_{\alpha}\}$ connected to the densities of the $\alpha$th species and fields $\{\omega_{\alpha}\}$ conjugate to $\{\rho_{\alpha}\}$. We study the correlations between these fields as well as their relations to the density correlations of the fluid.

The CV method is based on: (i) the concept of collective coordinates
being appropriate for the physics of the system considered (see, for
instance, \cite{Yuk-Hol}) and (ii) the integral identity
allowing to derive an exact functional representation for the
configurational Boltzmann factor. Being  applied to the continuous system the CV method uses the
idea of the reference system (RS), one of the basic ideas of the liquid state theory \cite{hansen_mcdonald}. The idea consists in the splitting of an interparticle interaction potential in two parts:
the potential of short-range repulsion which describes  the mutual impenetrability of the particles and the potential  describing mainly the behaviour at moderate and large distances. The equilibrium properties of the system interacting via the short-range repulsion are  assumed to be known. Therefore, this system can be regarded as the ``reference'' system. The remainder of the interaction is described in the phase space of CVs (collective coordinates). The fluid of hard spheres is most frequently used  as the RS in the liquid state theory
since its thermodynamic and structural properties are well known.
In this paper we derive the functional representation for the grand canonical partition function of a multicomponent fluid which includes both short-range and long-range interactions.

The paper is organized as follows. In Section ~2 we obtain the exact expression for the functional of the grand partition function of a multicomponent non-homogeneous mixture.  Section~3 is devoted to the study of the correlations of CVs fields and their relation to the density correlation functions of a multicomponent fluid. In Section~4 we formulate the perturbation theory. The MF level of the theory is considered in detail.

\section{The functional representation of the grand partition function}

\subsection{The model}
Let us consider a classical $m$-component system consisting of $N$ particles among which there exist $N_{1}$ particles of species $1$,  $N_{2}$ particles of species $2$, \ldots and  $N_{m}$ particles of species $m$.The potential energy of the system is assumed to be of the form
\begin{equation}
{\cal U}_{N_{1}\ldots N_{m}}=\frac{1}{2}\sum_{\alpha,\beta}^{m}\sum_{i\neq j}^{N}U_{\alpha\beta}({\mathbf r}_{i}^{\alpha},{\mathbf r}_{j}^{\beta})+\sum_{\alpha=1}^{m}\sum_{i=1}^{N}\psi_{\alpha}({\mathbf r}_{i}^{\alpha}),
\label{2.1}
\end{equation}
where $U_{\alpha\beta}({\mathbf r}_{i}^{\alpha},{\mathbf r}_{j}^{\beta})$ denotes the interaction potential of two particles and the second term is the potential energy due to external forces.

We present the pair interaction potential $U_{\alpha\beta}({\mathbf r}_{i}^{\alpha},{\mathbf r}_{j}^{\beta})$ as
\begin{equation}
U_{\alpha\beta}({\mathbf r}_{i}^{\alpha},{\mathbf r}_{j}^{\beta})=v_{\alpha\beta}^{0}({\mathbf r}_{i}^{\alpha},{\mathbf r}_{j}^{\beta})+w_{\alpha\beta}({\mathbf r}_{i}^{\alpha},{\mathbf r}_{j}^{\beta}),
\label{split}
\end{equation}
where $v_{\alpha\beta}^{0}({\mathbf r}_{i}^{\alpha},{\mathbf r}_{j}^{\beta})$ is a potential of a short-range repulsion
that can be chosen as an interaction between two hard spheres
of respective diameters $\sigma_{\alpha}$ and  $\sigma_{\beta}$.
We call
the $m$-component system with the interaction $v_{\alpha\beta}^{0}({\mathbf r}_{i}^{\alpha},{\mathbf r}_{j}^{\beta})$ a  reference  system  (RS). The thermodynamic and structural properties of the RS are assumed to be known.
$w_{\alpha\beta}({\mathbf r}_{i}^{\alpha},{\mathbf r}_{j}^{\beta})$ is some potential which can describe both repulsion (e.d. soft repulsion) and attractive interactions. In general,$w_{\alpha\beta}({\mathbf r}_{i}^{\alpha},{\mathbf r}_{j}^{\beta})$ can be presented in the form
\[
w_{\alpha\beta}({\mathbf r}_{i}^{\alpha},{\mathbf r}_{j}^{\beta})=w_{\alpha\beta}^{R}({\mathbf r}_{i}^{\alpha},{\mathbf r}_{j}^{\beta})+w_{\alpha\beta}^{A}({\mathbf r}_{i}^{\alpha},{\mathbf r}_{j}^{\beta}),
\]
where $w_{\alpha\beta}^{R}({\mathbf r}_{i}^{\alpha},{\mathbf
r}_{j}^{\beta})$ and $w_{\alpha\beta}^{A}({\mathbf
r}_{i}^{\alpha},{\mathbf r}_{j}^{\beta})$ are repulsive and
attractive parts of the interaction potential
$w_{\alpha\beta}({\mathbf r}_{i}^{\alpha},{\mathbf
r}_{j}^{\beta})$. Since $w_{\alpha\beta}({\mathbf
r}_{i}^{\alpha},{\mathbf r}_{j}^{\beta})$ are arbitrary in the
core, i.e.for $r\leq\sigma_{\alpha\beta}\equiv (\sigma_{\alpha}+\sigma_{\beta})/2 $, we assume that the
$w_{\alpha\beta}({\mathbf r}_{i}^{\alpha},{\mathbf
r}_{j}^{\beta})$ have been regularized in such a way that their
Fourier transforms  $\widetilde w_{\alpha\beta}(k)$ are
well-behaved functions of $k_{i}$ and that $w_{\alpha\beta}(0)$
are finite quantities. We denote by $\Omega$ the domain of volume
$V$ occupied by particles.

We present the potential energy ${\cal U}_{N_{1}\ldots N_{m}}$ as follows
\begin{equation}
{\cal U}_{N_{1}\ldots N_{m}}={\cal V}^{RS}_{N_{1}\ldots N_{m}}
+\frac{1}{2}\langle\widehat\rho_{\alpha}\vert w_{\alpha\beta}\vert\widehat\rho_{\beta}\rangle
+\langle\psi_{\alpha}\vert\widehat\rho_{\alpha}\rangle-N_{\alpha}\nu_{\alpha}^{S},
\label{energy}
\end{equation}
where
\begin{equation}
\widehat\rho_{\alpha}({\mathbf r})=\sum_{i=1}^{N_{\alpha}}\delta({\mathbf r}-{\mathbf r}_{i}^{\alpha})
\label{density}
\end{equation}
is the microscopic density of the $\alpha$th species in a given configuration and
$\psi_{\alpha}({\mathbf r})$ is some external one-body potential acting on particles of species $\alpha$.
The following notations are introduced in (\ref{energy}): ${\cal V}^{RS}_{N_{1}\ldots N_{m}}$
is the contribution from a $m$-component RS, $\nu_{\alpha}^{S}$ is the self-energy of the $\alpha$th species
\begin{equation}
\nu_{\alpha}^{S}=\frac{1}{2}w_{\alpha\alpha}(0).
\label{self_energy}
\end{equation}
In (\ref{energy}) we have also introduced Dirac's brackets notations
\begin{eqnarray*}
\langle\widehat\rho_{\alpha}\vert w_{\alpha\beta}\vert\widehat\rho_{\beta}\rangle=
\int_{\Omega} {\rm d}{\mathbf r}_{1}^{\alpha}{\rm d}{\mathbf r}_{2}^{\beta}\; \;
\widehat\rho_{\alpha}({\mathbf r}_{1}^{\alpha})w_{\alpha\beta}({\mathbf r}_{1}^{\alpha},
{\mathbf r}_{2}^{\beta})\widehat\rho_{\beta}({\mathbf r}_{2}^{\beta}),
\end{eqnarray*}
\begin{eqnarray*}
\langle\psi_{\alpha}\vert\widehat\rho_{\alpha}\rangle=\int_{\Omega}
{\rm d}{\mathbf r}_{1}^{\alpha} \;\psi_{\alpha}({\mathbf r}_{1}^{\alpha})\widehat\rho_{\alpha}({\mathbf r}_{1}^{\alpha}).
\end{eqnarray*}
In the above formulas summation over repeated indices is meant.

The system under consideration is at equilibrium in the grand
canonical (GC) ensemble, $\beta=1/k_{B}T$ is the inverse
temperature ($k_{B}$ Boltzmann constant), $\mu_{\alpha}$ is the
chemical potential of the $\alpha$th species. Then, the GC
partition function can be written as
\begin{eqnarray}
\Xi[\{\nu_{\alpha}\}]&=&\sum_{N_{1}\geq 0}\frac{1}{N_{1}!}
\sum_{N_{2}\geq 0}\frac{1}{N_{2}!}\ldots\sum_{N_{m}\geq 0}\frac{1}{N_{m}!}
\int({\rm d}\Gamma)\exp\left[-\beta{\cal V}^{\mathrm{RS}}_{N_{1}\ldots N_{m}}\right.
\nonumber \\
&&
\left.  -\frac{\beta}{2}\langle\widehat\rho_{\alpha}\vert w_{\alpha\beta}\vert\widehat\rho_{\beta}\rangle
+\langle\overline\nu_{\alpha}\vert\widehat\rho_{\alpha}\rangle\right],
\label{2.4}
\end{eqnarray}
where $\overline\nu_{\alpha}({\mathbf
r})=\nu_{\alpha}+\nu_{\alpha}^{S}-\beta\psi_{\alpha}({\mathbf r})$
is the local chemical potential of the $\alpha$th species and
$\nu_{\alpha}=\beta\mu_{\alpha}-3\ln\Lambda_{\alpha}$,
$\Lambda_{\alpha}^{-1}=(2\pi m_{\alpha}\beta^{-1}/h^{2})^{1/2}$ is
the inverse de Broglie thermal wavelength. $(\rm
d\Gamma)=\prod_{\alpha}{\rm d}\Gamma_{N_{\alpha}}$, ${\rm
d}\Gamma_{N_{\alpha}}={\rm d}{\mathbf r}_{1}^{\alpha}{\rm d}{\mathbf
r}_{2}^{\alpha}\ldots{\rm d}{\mathbf r}_{N_{\alpha}}^{\alpha}$ is
the element of the configurational space of $N$ particles.

For a given volume $V$, $\Xi[{\nu_{\alpha}}]$ is a function of the
temperature $T$ and a log-convex functional of the local chemical potentials $\nu_{\alpha}({\mathbf r}^{\alpha})$.

\subsection{The collective variables representation}

We introduce the collective variable  $\rho_{\alpha}({\mathbf r})$
which describes the field of the number particle density of the $\alpha$th species. To this end we use the identity
\begin{eqnarray}
\label{a}
\exp \left(
\frac{1}{2}\left\langle \widehat{\rho}\vert w \vert \widehat{\rho}\right\rangle
\right) = \int \mathcal{D} \rho \;
 \delta_{\mathcal{F}}\left[ \rho -\widehat{\rho} \right]
 \exp \left(
\frac{1}{2}\left\langle \rho \vert w \vert \rho \right\rangle
\right).
\end{eqnarray}
In (\ref{a}) the functional ``delta function'' $\delta_{\mathcal{F}}\left[ \rho \right]$ is defined as \cite{Orland}
\begin{equation}
\label{deltaF}
\delta_{\mathcal{F}}\left[ \rho\right] \equiv
 \int \mathcal{D} \omega \; \exp \left(i \left\langle \omega
 \vert \rho  \right\rangle \right) \;,
\end{equation}
Using (\ref{deltaF}) we can present the Boltzmann factor which does not include the RS interaction in the form
\begin{eqnarray}
\label{b}
\exp \left(
\frac{1}{2}\left\langle \widehat{\rho}_{\alpha}\vert w_{\alpha\beta} \vert \widehat{\rho_{\beta}}\right\rangle
\right)& =& \int \mathcal{D} \rho  \mathcal{D} \omega \;
\exp \left(  \frac{1}{2}\left\langle \rho_{\alpha} \vert w_{\alpha\beta} \vert \rho_{\beta} \right\rangle\right.
 \nonumber \\
&&
\left.  +{\rm i} \left\langle \omega_{\alpha} \vert \left\lbrace
 \rho_{\alpha} - \widehat{\rho}_{\alpha}
  \right\rbrace \right\rangle
 \right).
\end{eqnarray}
Inserting equation (\ref{b}) in the definition (\ref{2.4}) of the GC partition function $\Xi[{\nu_{\alpha}}]$ one obtains
\begin{equation}
\label{csiCV_1} \Xi\left[{ \{\nu_{\alpha}\} }\right]= \int
\mathcal{D} \rho \; \exp\left(  -\frac{\beta}{2}\left\langle
\rho_{\alpha} \vert w_{\alpha\beta} \vert \rho_{\beta}
\right\rangle\right) {\mathcal J}[\{\rho_{\alpha},\overline\nu_{\alpha}\}]
 \; ,
\end{equation}
where the Jacobian
\begin{equation}
\label{jacobian} {\mathcal J}[\{\rho_{\alpha},\overline\nu_{\alpha}\}]=\int
\mathcal{D}\omega\; \exp \left({\rm
i}\langle\omega_{\alpha}\vert\rho_{\alpha}\rangle\right)
\Xi_{RS}[\{\overline\nu_{\alpha}-{\rm i}\omega_{\alpha}\}] \;
\end{equation}
allows one for the passage from the microscopic variables ${\mathbf r}^{\alpha}$ (the Cartesian coordinates of particles) to the collective variables $\rho_{\alpha}({\mathbf r})$ (fields of the number density of particles). In (\ref{jacobian}) $\Xi_{\mathrm{RS}}[\{\overline\nu_{\alpha}-{\rm i}\omega_{\alpha}\}]=\Xi_{\mathrm{RS}}[Z^{*}]$ is the GC partition function of a $m$-component RS
\begin{eqnarray}
\Xi_{\mathrm{RS}}[Z^{*}]&=&\sum_{N_{1}\geq 0}\frac{1}{N_{1}!}\sum_{N_{2}\geq 0}\frac{1}{N_{2}!}\ldots\sum_{N_{m}\geq 0}\frac{1}{N_{m}!}\int({\rm d}\Gamma)
\nonumber \\
&&\times
\exp\left(-\beta {\cal V}^{\mathrm{RS}}_{N_{1}\ldots N_{m}}\right)\prod_{i}Z^{*}({\mathbf r}_{i}),
\label{2.11a}
\end{eqnarray}
where $Z^{*}({\mathbf r})=\prod_{\alpha}Z_{\alpha}^{*}({\mathbf
r})=\exp(\nu_{\alpha}^{*}({\mathbf r}))$.  $Z_{\alpha}^{*}$ is the activity of the
species $\alpha$ associated with the dimensionless local chemical potential
$\nu_{\alpha}^{*}({\mathbf r})=\overline\nu_{\alpha}(r)-{\rm
i}\omega_{\alpha}({\mathbf r})$. It should be noted that
${\mathcal J}[\{\rho_{\alpha},\overline\nu_{\alpha}\}]$ does not depend on
the pair interaction $w_{\alpha\beta}({\mathbf r}_{i},{\mathbf
r}_{j})$ but only on the GC partition function of the RS
$\Xi_{\mathrm{RS}}[Z^{*}]$ which is supposed to be known.

Equation (\ref{csiCV_1}) can also easily be recast in the form of a standard statistical field theory, i.e. as
\begin{equation}
\Xi\left[\{ {\nu_{\alpha}} \}\right]=\int \mathcal{D} \rho
\mathcal{D} \omega\; \exp \left(- {\mathcal
H}[\{\nu_{\alpha},\rho_{\alpha},\omega_{\alpha}\}]\right) \;,
\label{VSS}
\end{equation}
where the action ${\mathcal H}[\{\nu_{\alpha},\rho_{\alpha},\omega_{\alpha}\}]$ of the CV field theory reads as
\begin{equation}
\label{actionCV} 
\mathcal{H} \left[\{\nu_{\alpha}, \rho_{\alpha},
\omega_{\alpha}\} \right]=\frac{\beta}{2} \left\langle \rho_{\alpha}
\vert w_{\alpha\beta}\vert \rho_{\beta} \right\rangle  -{\rm i}
\left\langle \omega_{\alpha} \vert \rho_{\alpha}\right\rangle -
\ln\Xi_{\mathrm{RS}}\left[\{ \overline{\nu}_{\alpha} -{\rm i}
\omega_{\alpha}\} \right] \; .
\end{equation}

Functional integrals which enter the above-mentioned formulas can
be given a precise meaning in the case where the domain  $\Omega$
is a cube of side $L$ ( $V=L^{3}$) with periodic boundary
conditions. This means that we restrict ourselves to fields
$\rho_{\alpha}({\mathbf r})$ and $\omega_{\alpha}({\mathbf r})$
which can be written as Fourier series
\begin{equation}
\rho_{\alpha}(\mathbf{r})=\frac{1}{L^3} \; \sum_{\mathbf{k} \in
\Lambda} \rho_{\mathbf{k},\alpha} \; e^{{\rm i}
\mathbf{k}\mathbf{r}} \;,
\end{equation}
and
\begin{equation}
\omega_{\alpha}(\mathbf{r})=\frac{1}{L^3} \; \sum_{\mathbf{k} \in
\Lambda} \omega_{\mathbf{k},\alpha} \; e^{{\rm i}
\mathbf{k}\mathbf{r}} \;,
\end{equation}
where $\Lambda=(2\pi/L)\;  {\mathbb Z}^3$ is the reciprocal cubic lattice. The
reality of $\rho_{\alpha}$ (and $\omega_{\alpha}$) implies that,
for ${\mathbf k}\neq 0$ $\rho_{-{\mathbf k},\alpha}=\rho_{{\mathbf
k},\alpha}^{\star}$ ($\omega_{-{\mathbf
k},\alpha}=\omega_{{\mathbf k},\alpha}^{\star}$), where the star
means complex conjugation. Then, the normalized functional measure
$\mathcal{D}\rho$ (and $\mathcal{D}\omega$) is defined as \cite{Wegner}

\begin{eqnarray}
\label{dphi}
\mathcal{D} \rho & \equiv & \prod_{\alpha}\prod_{\mathbf{k} \in
\Lambda} \frac{{\rm d} \rho_{\mathbf{k},\alpha} }
{\sqrt{2 \pi  V}} \\
{\rm d} \rho_{\mathbf{k},\alpha} {\rm d} \rho_{-\mathbf{k},\alpha}
& = & 2 \; {\rm d}\Re{\rho_{\mathbf{k},\alpha}} \; {\rm
d}\Im{\rho_{\mathbf{k},\alpha}},  \qquad \mathbf{k } \ne 0 \; .
\end{eqnarray}
Equation (\ref{dphi}) can be rewritten as
\begin{equation}
\label{dphi_bis} \mathcal{D} \rho= \prod_{\alpha}\frac{{\rm d}
\rho_{0,\alpha}}{\sqrt{2 \pi  V}} \prod_{\mathbf{q}
\in \Lambda^{\star}} \frac{{\rm d}\Re{\rho_{\mathbf{q},\alpha}} \;
{\rm d}\Im{\rho_{\mathbf{q},\alpha}} }{\pi V} \; ,
\end{equation}
where the sum in the r.h.s. runs over only the half $\Lambda^{\star}$ of all the vectors of the reciprocal lattice $\Lambda$. We have for $\mathcal{D}\omega$, respectively
\begin{equation}
\label{domega_bis} 
\mathcal{D} \omega= \prod_{\alpha}\frac{{\rm d}
\omega_{0,\alpha}}{\sqrt{2 \pi  V}}
\prod_{\mathbf{q} \in \Lambda^{\star}} \frac{{\rm
d}\Re{\omega_{\mathbf{q},\alpha}} \; {\rm
d}\Im{\omega_{\mathbf{q},\alpha}} }{\pi V} \; .
\end{equation}

Now let us present the action (\ref{actionCV}) for the isotropic interaction potential $w_{\alpha\beta}(r)$ as follows
\begin{eqnarray}
{\cal
H}[\{\nu_{\alpha},\rho_{\alpha},\omega_{\alpha}\}]&=&\frac{1}{2}\sum_{\alpha,\beta}\sum_{{\mathbf
k}}\widetilde \Phi_{\alpha\beta}(k)\rho_{{\mathbf
k},\alpha}\rho_{-{\mathbf k},\beta}-{\rm
i}\sum_{\alpha}\sum_{{\mathbf k}}\omega_{{\mathbf
k},\alpha}\rho_{{\mathbf k},\alpha}
\nonumber\\
&& -\ln \Xi_{\mathrm{RS}}[\{\bar \nu_{\alpha}-{\rm i}\omega_{\alpha}\}],
\label{actionCV_rhok}
\end{eqnarray}
Here CV $\rho_{{\mathbf k},\alpha}$ describes the $k$th mode of number density fluctuations of the $\alpha$th species. $\widetilde \Phi_{\alpha\beta}(k)=\frac{\beta}{V}\widetilde w_{\alpha\beta}(k)$, where $\widetilde w_{\alpha\beta}(k)$ is the Fourier transform of the interaction potential $w_{\alpha\beta}(r)$.

In order to obtain another equivalent representation of the action ${\cal H}[\{\nu_{\alpha},\rho_{\alpha},\omega_{\alpha}\}]$ we first distinguish
the chemical potential $\nu_{\alpha}^{0}$ of the particle of the  species $\alpha$ in the RS.  To this end we put
\begin{equation}
\overline\nu_{\alpha}-{\rm i}\omega_{\alpha}({\mathbf
r})=\nu_{\alpha}^{0}-{\rm i}\omega_{\alpha}^{'}({\mathbf r})
\label{chem_shift}
\end{equation}
and obtain
\[
{\rm i}\omega_{\alpha}({\mathbf r})=\Delta\nu_{\alpha}+{\rm
i}\omega_{\alpha}^{'}({\mathbf r})
\]
with $\Delta\nu_{\alpha}=\overline\nu_{\alpha}-\nu_{\alpha}^{0}$.
As a result, the action (or the Hamiltonian) (\ref{actionCV_rhok}) can be written as
\begin{eqnarray}
{\cal
H}[\{\nu_{\alpha},\rho_{\alpha},\omega_{\alpha}\}]&=&-\sum_{\alpha}\Delta\nu_{\alpha}\rho_{{\mathbf
k=0},\alpha}+\frac{1}{2}\sum_{\alpha,\beta}\sum_{{\mathbf
k}}\widetilde \Phi_{\alpha\beta}(k)\rho_{{\mathbf
k},\alpha}\rho_{-{\mathbf k},\beta}
\nonumber\\
&& -{\rm i}\sum_{\alpha}\sum_{{\mathbf k}}\omega_{{\mathbf
k},\alpha}^{'}\rho_{{\mathbf k},\alpha}-\ln
\Xi_{\mathrm{RS}}[\{\nu_{\alpha}^{0}-{\rm i}\omega_{\alpha}^{'}\}].
\label{actionCV_rhok1}
\end{eqnarray}

We have obtained the exact representations of the grand partition function of a multicomponent system (eqs. (\ref{VSS})-(\ref{actionCV}), (\ref{actionCV_rhok}) and (\ref{actionCV_rhok1})) in terms of CVs $\rho_{\alpha}(\mathbf{r})$, which are number density fields of the $\alpha$th species particles (or fluctuation modes of the $\alpha$th species number density) We also stress that $\rho_{\alpha}({\mathbf r})$ and $\omega_{\alpha}({\mathbf r})$ are two real scalar fields  and that eqs.~(\ref{VSS})-(\ref{actionCV}) (as well as eqs.~(\ref{actionCV_rhok}) and
(\ref{actionCV_rhok1})) are valid for
repulsive, attractive as well as  arbitrary pair interactions.

Besides the usual GC average $\langle{\mathcal
A}\rangle_{\mathrm{GC}}$ of a dynamic variable we introduce statistical field averages of the type
\begin{equation}
\label{moyCV} 
\left\langle \mathcal{A}\left[\{\rho_{\alpha},
\omega_{\alpha}\} \right]
\right\rangle_{\mathrm{CV}}=\Xi\left[\{\nu_{\alpha}\} \right]^{-1} \;
\int \mathcal{D} \rho
\mathcal{D} \omega \; \; \mathcal{A}\left[\{\rho_{\alpha}, \omega_{\alpha}\} \right]
\exp \left( - \mathcal{H}\left[\{\nu_{\alpha}, \rho_{\alpha},
\omega_{\alpha}\} \right] \right),
\end{equation}
where $\mathcal{A}\left[\{\rho_{\alpha}, \omega_{\alpha}\}\right]$ is a functional of the two CV fields $\rho_{\alpha}$ and $\omega_{\alpha} $.

\section{Correlation functions}
\subsection{General relations}

Let us write some important relations. First, according to \cite{stell, stell1} we introduce the ordinary and truncated (or connected) density correlation functions
\begin{eqnarray}
\label{defcorre}
G_{\alpha_{1}\ldots\alpha_{n}}^{(n)}[\{\nu_{\alpha}\}](1, \ldots, n)
&=&\left< \prod_{1=1}^{n} \widehat{\rho}_{\alpha_{i}}
    (i)  \right>_{\mathrm{GC}}
\nonumber \\
&=&\frac{1}{\Xi[{\nu_{\alpha}}]}\frac{\delta^{n}
\;\Xi[\{\nu_{\alpha}\}]}
{\delta \nu_{\alpha_{1}}(1) \ldots \delta \nu_{\alpha_{n}}(n)}           \; ,\nonumber \\
G_{\alpha_{1}\ldots\alpha_{n}}^{(n), T}[\{\nu_{\alpha}\}](1, \ldots,
n) &=&  \frac{\delta^{n} \log \Xi[\{\nu_{\alpha}\}]} {\delta
\nu_{\alpha_{1}}(1) \ldots \delta \nu_{\alpha_{n}}(n)} \; .
\end{eqnarray}
Our notation emphasizes the fact that the correlation functions
(connected and not connected) are functionals of the local chemical potential
 $\nu_{\alpha}({\mathbf r})$  and
functions of the coordinates $(1,2,\ldots,n)\equiv ({\mathbf
r}_{1},{\mathbf r}_{2},\ldots,{\mathbf r}_{n})$. For the sake of simplicity, we omit below the notations which indicate the functional dependence of the correlation functions of $\nu_{\alpha}({\mathbf r})$.
In standard textbooks of liquid theory \cite{hansen_mcdonald} the $n$-particle correlation functions are more frequently defined as functional derivatives of $\Xi$ or $\log\Xi$ with respect to the activities $Z_{\alpha}=\exp(\nu_{\alpha})$ rather than with respect to the local chemical potentials
\begin{equation}
\frac{\rho_{\alpha_{1}\ldots\alpha_{n}}(1,2,\ldots,n)}{Z_{\alpha_{1}}^{*}(1)Z_{\alpha_{2}}^{*}(2)\ldots
Z_{\alpha_{n}}^{*}(n)}=\frac{1}{\Xi}\frac{\delta^{n}\Xi}{\delta
Z_{\alpha_{1}}^{*}(1)\delta Z_{\alpha_{2}}^{*}(2)\ldots\delta
Z_{\alpha_{n}}^{*}(n)}, \label{3.5}
\end{equation}
\begin{equation}
\frac{\rho_{\alpha_{1}\ldots\alpha_{n}}^{T}(1,2,\ldots,n)}{Z_{\alpha_{1}}^{*}(1)Z_{\alpha_{2}}^{*}(2)\ldots
Z_{\alpha_{n}}^{*}(n)}=\frac{1}{\Xi}\frac{\delta^{n}\ln\Xi}{\delta
Z_{\alpha_{1}}^{*}(1)\delta Z_{\alpha_{2}}^{*}(2)\ldots\delta
Z_{\alpha_{n}}^{*}(n)}. \label{3.5a}
\end{equation}

We also define the partial distribution function $g_{\alpha_{1}\ldots\alpha_{n}}(1,2,\ldots,n)$ and the partial correlation functionі  $h_{\alpha_{1}\ldots\alpha_{n}}(1,2,\ldots,n)$ \cite{stell}
\begin{equation}
g_{\alpha_{1}\ldots\alpha_{n}}(1,2,\ldots,n)=\frac{\rho_{\alpha_{1}\ldots\alpha_{n}}(1,2,\ldots,n)}{\rho_{\alpha_{1}}(1)\rho_{\alpha_{2}}(2)\ldots\rho_{\alpha_{n}}(n)}
\label{3.5c}
\end{equation}
and
\begin{equation}
h_{\alpha_{1}\ldots\alpha_{n}}(1,2,\ldots,n)=\frac{\rho_{\alpha_{1}\ldots\alpha_{n}}^{T}(1,2,\ldots,n)}{\rho_{\alpha_{1}}(1)\rho_{\alpha_{2}}(2)\ldots\rho_{\alpha_{n}}(n)}.
\label{3.5b}
\end{equation}

Expressions (\ref{defcorre}) and (\ref{3.5c})-(\ref{3.5b}) differ by the terms involving products of delta functions. For instance, for $n=2$ and for a homogeneous system one has
\begin{eqnarray}
G_{\alpha\beta}^{(2)}(1,2) &=&\rho_{\alpha}\rho_{\beta}g_{\alpha\beta}(1,2)+\rho_{\alpha}\delta_{\alpha\beta}\delta(1,2),\nonumber\\
G_{\alpha\beta}^{(2), T}(1,2)
&=&\rho_{\alpha}\rho_{\beta}h_{\alpha\beta}(1,2)+\rho_{\alpha}\delta_{\alpha\beta}\delta(1,2),
\label{correl_2}
\end{eqnarray}
where  $\rho_{\alpha}$ is the equilibrium density of the species $\alpha$. Besides,
functions  $h_{\alpha\beta}(r)$ and  $g_{\alpha\beta}(r)$ are connected by the relation
$h_{\alpha\beta}(r)=g_{\alpha\beta}(r)-1$ for a homogeneous case.

\subsection{CV field correlations}

Let us consider the correlations of fields $\rho_{\alpha}$ and $\omega_{\alpha}$. We start with the definitions of the correlation functions
\begin{eqnarray}
\label{G-CV}
G^{(n)}_{\rho_{\alpha_{1}}\ldots\rho_{\alpha_{n}}}[\{\nu_{\alpha}\}](1,
\ldots, n) &=&\left< \prod_{i=1}^{n} \rho_{\alpha_{i}} \left(i
\right)\right>_{\mathrm{CV}} \; , \nonumber \\
G^{(n)}_{\omega_{\alpha_{1}}\ldots\omega_{\alpha_{n}}}[\{\nu_{\alpha}\}](1,
\ldots, n) &=&\left< \prod_{i=1}^{n} \omega_{\alpha_{i}} \left(i
\right)\right>_{\mathrm{CV}} \; ,
\end{eqnarray}
and their truncated (connected) parts
\begin{eqnarray}
\label{G-CV2} G^{(n),
T}_{\rho_{\alpha_{1}}\ldots\rho_{\alpha_{n}}}(1,\ldots,n)&= &
G^{(n)}_{\rho_{\alpha_{1}}\ldots\rho_{\alpha_{n}}}( 1,\ldots,n)
\nonumber \\
&& -\sum \prod_{m<n}G^{(m),
T}_{\rho_{\alpha_{1}}\ldots\rho_{\alpha_{m}}}(i_{1},\ldots,i_{m})
\; ,
\nonumber \\
G^{(n), T}_{\omega_{\alpha_{1}}\ldots\omega_{\alpha_{n}}}
(1,\ldots,n)&= &
G^{(n)}_{\omega_{\alpha_{1}}\ldots\omega_{\alpha_{n}}}(
1,\ldots,n)
\nonumber \\
&& - \sum \prod_{m<n}G^{(m),
T}_{\omega_{\alpha_{1}}\ldots\omega_{\alpha_{m}}}(i_{1},\ldots,i_{m})
\; , \;
\end{eqnarray}
where the sum of products is carried out over all possible partitions of the set $(1,\ldots,n)$ into subsets of cardinality $m<n$.

\paragraph{Correlation functions $G^{(n)}_{\rho_{\alpha_{1}}\ldots\rho_{\alpha_{n}}}$.}

Now we introduce the modified partition function
\begin{equation}
\label{Xi1} \Xi^1\left[\{\nu_{\alpha},J_{\alpha}\} \right]=
 \int \mathcal{D} \rho
 \mathcal{D} \omega \;
\exp \left( - \mathcal{H}_{\mathrm{CV}}\left[\{\nu_{\alpha},
\rho_{\alpha}, \omega_{\alpha}\} \right]+\left\langle  J_{\alpha}
\vert \rho_{\alpha}\right\rangle \right) \; ,
\end{equation}
where $J_{\alpha}$ is a real scalar field and 
$\Xi^1\left[\{\nu_{\alpha},J_{\alpha}\}\right]$ is the generator of field correlation functions $G^{(n)}_{\rho_{\alpha_{1}}\ldots\rho_{\alpha_{n}}}$ \cite{Zinn}. As a result, we have
\begin{eqnarray}
\label{defcorreCVrho}
G^{(n)}_{\rho_{\alpha_{1}}\ldots\rho_{\alpha_{n}}}[\{\nu_{\alpha}\}](1,
\ldots, n) &=& \frac{1}{\Xi^1[\{\nu_{\alpha}\}]} \left.
\frac{\delta^{n} \;\Xi^1[\{\nu_{\alpha},J_{\alpha}\}]}
{\delta J_{\alpha_{1}}(1) \ldots \delta J_{\alpha_{n}}(n)} \right \vert_{J_{\alpha_{i}}=0}          \; ,\nonumber \\
G^{(n), T}_
{\rho_{\alpha_{1}}\ldots\rho_{\alpha_{n}}}[\{\nu_{\alpha}\}](1,
\ldots, n) &=& \left.  \frac{\delta^{n} \log
\Xi^1[\{\nu_{\alpha},J_{\alpha}\}]} {\delta J_{\alpha_{1}}(1) \ldots
\delta J_{\alpha_{n}}(n)}  \right \vert_{J_{\alpha_{i}}=0}\; .
\end{eqnarray}
The simplest way to obtain the relations between the
$G^{(n)}_{\rho_{\alpha_{1}}\ldots\rho_{\alpha_{n}}}[\{\nu_{\alpha}\}](1,
\ldots, n)$ and the density correlation functions is to start
from the definition~(\ref{defcorre}). One has
\begin{eqnarray}
G^{(n)}_{\alpha_{1}\ldots\alpha_{n}}(1, \ldots, n)&=&
\frac{1}{\Xi[\{\nu_{\alpha}\}]}\frac{\delta^{n} \;\Xi[\{\nu_{\alpha}\}]}
{\delta \nu_{\alpha_{1}}(1) \ldots \delta \nu_{\alpha_{n}}(n)}
\;
\nonumber \\
&=&\frac{1}{\Xi[\{\nu_{\alpha}\}]}
\int \mathcal{D} \rho
\mathcal{D} \omega \;
\exp \left(\frac{1}{2} \left\langle\rho_{\alpha} \vert w_{\alpha\beta}^{*}\vert \rho_{\beta} \right\rangle\right.
\nonumber
 \\
 && 
\left.  +{\rm i} \left\langle \omega_{\alpha}\vert \rho_{\alpha} \right\rangle \right)
\frac{\delta^{n} \; \Xi_{\mathrm{RS}}[\{\overline{\nu}_{\alpha}
-{\rm i} \omega_{\alpha}\}]} {\delta \nu_{\alpha_{1}}(1) \ldots
\delta\nu_{\alpha_{n}}(n)} \nonumber
 \\
&=&\frac{1}{\Xi[\{\nu_{\alpha}\}]}
\int \mathcal{D} \rho
\mathcal{D} \omega \;
\exp \left(\frac{1}{2} \left\langle\rho_{\alpha} \vert w_{\alpha\beta}^{*}\vert \rho_{\beta} \right\rangle\right. 
\nonumber
\\
&&\left. +{\rm i} \left\langle \omega_{\alpha}\vert \rho_{\alpha} \right\rangle \right)  \nonumber \\
&\times& ({\rm i})^n
\frac{\delta^{n} \;
\Xi_{\mathrm{RS}}[\{\overline{\nu} _{\alpha} -{\rm i} \omega_{\alpha}\}]}
{\delta \omega_{\alpha_{1}}(1) \ldots \delta
\omega_{\alpha_{n}}(n)} \nonumber \; ,
\end{eqnarray}
where we introduce the notation $w_{\alpha\beta}^{*}=-\beta w_{\alpha\beta}$
and use the equality
\begin{equation}
\frac{\delta^{n} \; \Xi_{\mathrm{RS}}[\{\overline{\nu}_{\alpha} -{\rm
i} \omega_{\alpha}\}]} {\delta \nu_{\alpha_{1}}(1) \ldots \delta
\nu_{\alpha_{n}}(n)} = ({\rm i})^n
 \frac{\delta^{n} \;
 \Xi_{\mathrm{RS}}[\{\overline{\nu} _{\alpha} -{\rm i} \omega_{\alpha}\}]}
{\delta \omega_{\alpha_{1}}(1) \ldots \delta
\omega_{\alpha_{n}}(n)}. \label{eqality}
\end{equation}
Performing now $n$ integral by parts yields
\begin{eqnarray}
G^{(n)}_{\alpha_{1}\ldots\alpha_{n}}(1, \ldots, n)&=&\frac{(-{\rm
i})^n}{\Xi[\{\nu_{\alpha}\}]} \int \mathcal{D} \rho \mathcal{D} \omega
\;
\exp \left(\frac{1}{2} \left\langle\rho_{\alpha} \vert w_{\alpha\beta}^{*}\vert \rho_{\beta} \right\rangle\right.
\nonumber \\
&&
\left.+ \ln \Xi_{\mathrm{RS}}[\{\overline{\nu}_{\alpha} -{\rm i} \omega_{\alpha}\}] \right)
\frac{\delta^{n} \;
\exp\left({\rm i}\left\langle \omega_{\alpha} \vert \rho_{\alpha} \right\rangle \right) }
{\delta \omega_{\alpha_{1}}(1) \ldots \delta
\omega_{\alpha_{n}}(n)}
\nonumber \\
&&
=\left< \prod_{i=1}^{n} \rho_{\alpha_{i}} \left(i\right) \right>_{\mathrm{CV}}
  \; . \nonumber
\end{eqnarray}
We have just proved the expected result
\begin{equation}
G^{(n)}_{\alpha_{1}\ldots\alpha_{n}}\left[\{\nu_{\alpha}\} \right] (1,
\ldots,
n)=G^{(n)}_{\rho_{\alpha_{1}}\ldots\rho_{\alpha_{n}}}\left[\{\nu_{\alpha}\}
\right] (1, \ldots, n) \; . \label{corelat}
\end{equation}
Obviously, the following relation is also valid for the truncated (connected) correlation functions
\begin{equation}
\label{dens-CV-rho} G^{(n),
T}_{\alpha_{1}\ldots\alpha_{n}}\left[\{\nu_{\alpha}\} \right] (1,
\ldots,
n)=G^{(n),T}_{\rho_{\alpha_{1}}\ldots\rho_{\alpha_{n}}}\left[\{\nu_{\alpha}\}
\right] (1, \ldots, n) \; .
\end{equation}

\paragraph{Correlation functions $G^{(n)}_{\omega_{\alpha_{1}}\ldots\omega_{\alpha_{n}}}$.}

Let us define the modified  partition function
\begin{equation}
\label{Xi3} \Xi^2\left[\{\nu_{\alpha},J_{\alpha}\}\right]=
 \int \mathcal{D} \rho
 \mathcal{D} \omega \;
\exp \left( - \mathcal{H}_{\mathrm{CV}} \left[\{\nu_{\alpha},
\rho_{\alpha}, \omega_{\alpha}\} \right]+\left\langle  J_{\alpha}
\vert \omega_{\alpha} \right\rangle \right) \; ,
\end{equation}
where $J_{\alpha}$ is a real scalar field.
$\Xi^2\left[\{\nu_{\alpha},J_{\alpha}\} \right]$ is the generator of the functions 
$G^{(n)}_{\omega_{\alpha_{1}}\ldots\omega_{\alpha_{n}}}$ and we thus have
\begin{eqnarray}
\label{defcorreCVomega}
G^{(n)}_{\omega_{\alpha_{1}}\ldots\omega_{\alpha_{n}}}[\{\nu_{\alpha}\}](1,
\ldots, n) &=& \frac{1}{\Xi^{2}[\{\nu_{\alpha}]\}} \left.
\frac{\delta^{n} \;\Xi^2[\{\nu_{\alpha},J_{\alpha}\}]} {\delta
J_{\alpha_{1}}(1) \ldots \delta J_{\alpha_{n}}(n)} \right
\vert_{J_{\alpha_{i}}=0}          \; ,\nonumber
\\ G^{(n), T}_{\omega_{\alpha_{1}}\ldots\omega_{\alpha_{n}}}[\{\nu_{\alpha}\}](1, \ldots, n) &=& \left.
\frac{\delta^{n} \log \Xi^2[\{\nu_{\alpha},J_{\alpha}\}]} {\delta
J_{\alpha_{1}}(1) \ldots \delta J_{\alpha_{n}(}n)}  \right
\vert_{J_{\alpha_{i}}=0}\; .
\end{eqnarray}
In order to relate  the correlation functions $G^{(n)}_{\omega_{\alpha_{1}}\ldots\omega_{\alpha_{n}}}$ and
$G^{(n)}_{\alpha_{1}\ldots\alpha_{n}}(1, \ldots, n)$  we perform the change of variables $\rho_{\alpha} \rightarrow \rho_{\alpha} +{\rm i}
J_{\alpha}$ in eq.~(\ref{Xi3}). The functional Jacobian of the transformation is of course equal to unity and one obtains the relation
\begin{equation}
\label{jojo} \ln \Xi^2\left[\{\nu_{\alpha},J_{\alpha}\} \right]=
-\frac{1}{2} \left\langle J_{\alpha}\vert w_{\alpha\beta}^{*} \vert
J_{\beta}  \right\rangle + \ln\Xi^1\left[\{\nu_{\alpha}, {\rm i} w_{\alpha\beta}^{*}
\star J_{\beta}\} \right] \;,
\end{equation}
where the star $\star$ means space convolution and 
$\Xi^1$ is defined in~(\ref{Xi1}). The idea is to perform now $n$
successive functional derivatives   of both sides of
eq.~(\ref{jojo}) with respect to $J_{\alpha}$. Since it follows from the
expression~(\ref{defcorreCVrho}) that
\begin{eqnarray}
\left. \frac{\delta^{n} \log \Xi^1[\{\nu_{\alpha},{\rm i} w_{\alpha\beta}^{*} \star
J_{\beta}\}]} {\delta J_{\alpha_{1}}(1) \ldots \delta J_{\alpha_{n}}(n)}
\right \vert_{J_{\alpha_{i}}=0}&=&
{\rm i}^n w_{\alpha_{1}\alpha_{1'}}^{*}(1,1^{'}) \ldots w_{\alpha_{n}\alpha_{n'}}^{*}(n,n^{'}) \nonumber \\
&\times&  G^{(n), T}_{\rho_{\alpha_{1'}}\ldots\rho_{\alpha_{n'}}}
[\{\nu_{\alpha}\}](1^{'}, \ldots, n^{'}),\nonumber \\
\end{eqnarray}
one obtains

\begin{eqnarray*}
\left\langle \omega_{\alpha_{1} }(1)
\right\rangle_{\mathrm{CV}}&=&{\rm i} \; w_{\alpha_{1}\alpha_{1'
}}^{*}(1,1^{'}) \left\langle \rho_{\alpha_{1'}} (1^{'})
\right\rangle_{\mathrm{CV}} \; , \\ \label{dens-CV2}
G^{(2), T}_{\omega_{\alpha_{1}}\omega_{\alpha_{2}}}\left[\{\nu_{\alpha}\} \right] (1,2)     &=&- w_{\alpha_{1}\alpha_{2}}^{*}(1,2)-w_{\alpha_{1}\alpha_{1'}}^{*}(1,1^{'})  \nonumber \\
& \times &w_{\alpha_{2}\alpha_{2'}}^{*}(2,2^{'})
G^{(2), T}_{\rho_{\alpha_{1'}}\rho_{\alpha_{2'}}}\left[\{ \nu_{\alpha}\} \right] (1^{'},2^{'}) \; ,\nonumber  \\
G^{(n), T}_{\omega_{\alpha_{1}}\ldots\omega_{\alpha_{n}}}\left[\{\nu_{\alpha}\} \right] (1,\ldots,n)&=& {\rm i}^n \;  w_{\alpha_{1}\alpha_{1'}}^{*}(1,1^{'})
\ldots w_{\alpha_{n}\alpha_{n'}}^{*}(n,n^{'})  \nonumber \\
&\times &
G^{(n), T}_{\rho_{\alpha_{1'}}\ldots\rho_{\alpha_{n'}}}\left[\{\nu_{\alpha}\} \right] (1',\ldots,n'),\quad  n\geq 3.
\nonumber \\
\end{eqnarray*}

\section{The perturbation theory}

\subsection{Mean-field theory}

Let us consider the functional of the GC partition function (\ref{VSS}) with the action given by eq. (\ref{actionCV_rhok}) for the case of an isotropic interaction potential $w_{\alpha\beta}(r)$. At the MF level one has \cite{Zinn}
\begin{equation}
\Xi_{\mathrm{MF}}[\{\nu_{\alpha}\}]=\exp(-{\mathcal
H}[\{\nu_{\alpha},\overline\rho_{\alpha}, \overline\omega_{\alpha}\}]),
\label{2.14}
\end{equation}
where, for $\overline\rho_{\alpha}$ and $\overline\omega_{\alpha}$, the action is stationary, i.e.
\begin{equation}
\left.\frac{\delta \; \mathcal{H}\left[\{\nu_{\alpha},
\rho_{\alpha}, \omega_{\alpha}\}\right]}{\delta
\rho_{\alpha}}\right
\vert_{(\overline\rho_{\alpha},\overline\omega_{\alpha})} =\left.
\frac{\delta \; \mathcal{H}\left[\{ \nu_{\alpha}, \rho_{\alpha},
\omega_{\alpha}\} \right]}{\delta \omega_{\alpha}}\right
\vert_{(\overline\rho_{\alpha},\overline\omega_{\alpha})} =0 \; .
\label{statio-CV}
\end{equation}
Replacing the CV action by its expression~(\ref{actionCV_rhok}) in eq.~(\ref{statio-CV})
leads to  implicit equations for  $\overline\rho_{\alpha}$ and
$\overline\omega_{\alpha}$:
\begin{eqnarray}
\overline\rho_{\alpha}(1)&=&\overline\rho_{\alpha}^{\mathrm{MF}}(1)=\rho_{\alpha}^{{\mathrm RS}}[\{\bar \nu_{\alpha}-{\rm i}\overline\omega_{\alpha}\}](1), \nonumber \\
{\rm{i}}\overline\omega_{\alpha}(1)&=&
\Phi_{\alpha\beta}(1,2)\overline\rho_{\beta}(2),
\label{2.15a}
\end{eqnarray}
where $\rho_{\alpha}^{RS}[\{\bar \nu_{\alpha}-{\rm
i}\overline\omega_{\alpha}\}](i)$ denotes the $\alpha$th species number density of the RS fluid with the chemical potentials $\{\bar \nu_{\alpha}-{\rm
i}\overline\omega_{\alpha}\}$. For a homogeneous system (\ref{2.15a}) can be rewritten in the form
\begin{eqnarray}
\overline\rho_{\alpha}&=&\overline\rho_{\alpha}^{\mathrm {MF}}=\rho_{\alpha}^{{\mathrm RS}}[\{\bar \nu_{\alpha}-{\rm i}\overline\omega_{\alpha}\}], \nonumber \\
{\rm{i}}\overline\omega_{\alpha}&=& \overline\rho_{\beta}\tilde{\Phi}_{\alpha\beta}(0),
\label{2.15}
\end{eqnarray}
It follows from the stationary condition (\ref{statio-CV}) that the MF density is given by
\begin{equation}
\label{ro-MF}
\rho_{\alpha}^{\mathrm{MF}}\left[\{\nu_{\alpha}\} \right] (1)= \frac{\delta \ln \Xi_{\mathrm{MF}}\left[\{\nu_{\alpha}\} \right]}{\delta \nu_{\alpha}(1)}=
\rho_{\alpha}^{\mathrm{RS}} \left[\left\lbrace \overline\nu_{\alpha}-{\rm i}\overline\omega_{\alpha}\right\rbrace \right](1) \; ,
\end{equation}
and that the MF grand potential reads
\begin{equation}
\label{MF-gpot}
\ln \Xi_{\mathrm{MF}}\left[\{\nu_{\alpha}\} \right]=
\ln
\Xi_{\mathrm{RS}}\left[\{\overline{\nu}_{\alpha}-{\rm
i}\overline\omega_{\alpha}\} \right]+ \frac{\beta}{2}\left\langle
\rho_{\alpha}^{\mathrm{MF}}\vert w_{\alpha\beta}
\vert\rho_{\beta}^{\mathrm{MF}}\right\rangle \; .
\end{equation}
The MF Kohn-Scham free energy of a multicomponent system defined as the Legendre transform
\begin{equation}
\beta \mathcal{A}_{\mathrm{MF}}\left[\{\rho_{\alpha}\}\right]
=\sup_{\nu_{\alpha}}\left\lbrace
\left\langle \rho_{\alpha} \vert \nu_{\alpha}\right\rangle -\ln \Xi_{\mathrm{MF}}\left[\{\nu_{\alpha}\}\right]
\right\rbrace,
\end{equation}
has the following form in the MF approximation
\begin{equation}
\label{MF-A} 
\beta \mathcal{A}_{\mathrm{MF}}\left[\{\rho_{\alpha}\}
\right]  = \beta \mathcal{A}_{\mathrm{RS}}\left[\{\rho_{\alpha}\}
\right] +\frac{\beta}{2}\left\langle\rho_{\alpha} \vert
w_{\alpha\beta} \vert \rho_{\beta} \right\rangle -\frac{\beta}{2}
\int_{\Omega} {\rm d} {\mathbf r}\; w_{\alpha\alpha}(0)
\rho_{\alpha}({\mathbf r}) \; .
\end{equation}
Using the formulas
\begin{eqnarray*}
G_{{\mathrm{MF}},\alpha_{1}\alpha_{2}}^{(2),T}\left[\{\nu_{\alpha}\}
\right](1,2)&=& \frac{\delta^{2} \ln \Xi_{\mathrm{\mathrm{MF}}}\left[
\{\nu_{\alpha}\}\right]}
{\delta \nu_{\alpha_{1}}(1) \; \delta \nu_{\alpha_{2}}(2)} \; , \\
C_{{\mathrm{MF}},\alpha_{1}\alpha_{2}}^{(2)}\left[\{\rho_{\alpha}\} \right]
(1,2)&=& -\frac{\delta^{2} \beta
\mathcal{A}_{\mathrm{\mathrm{MF}}}\left[\{\rho_{\alpha}\}\right]} {\delta
\rho_{\alpha_{1}}(1) \; \delta \rho_{\alpha_{2}}(2)} \; ,
\end{eqnarray*}
where $(1,2,\ldots,n)\equiv ({\mathbf r}_{1},{\mathbf r}_{2},\ldots,
{\mathbf r}_{n})$, one can get the well-known expressions for the partial pair correlation and vertex (or direct correlation) functions in the MF approximation. $C_{{\mathrm{MF}},\alpha_{1}\alpha_{2}}^{(2)}\left[\{\rho_{\alpha}\} \right](1,2)$ is obtained readily from the expression (\ref{MF-A})
\[
C_{{\mathrm{MF}},\alpha\beta}^{(2)}(1,2)=-G_{{\mathrm{MF}},\alpha\beta}^{(2),T-1}(1,2)=
C_{{\mathrm{RS}},\alpha\beta}^{(2)}(1,2)-w_{\alpha\beta}(1,2),
\]
where $C_{{\mathrm{RS}},\alpha\beta}^{(2)}(1,2)$ means the exact two-point proper vertex of the RS fluid at the mean field density $\rho_{\alpha}^{{\mathrm{MF}}}$. The two-point vertex function  $C_{\mathrm{MF},\alpha\beta}^{(2)}$ is connected to the usual direct correlation function of the theory of liquids  $c_{{\mathrm{MF}},\alpha\beta}(1,2)$
\[
C_{\mathrm{MF},\alpha\beta}^{(2)}(1,2)=c_{\mathrm{MF},\alpha\beta}(1,2)-\frac{1}{\rho_{\alpha}(1)}\delta_{\alpha\beta}\delta(1,2).
\]
In order to calculate $G_{{\mathrm{MF}},\alpha\beta}^{(2),T}(1,2)$ we start with equation
\begin{equation}
G_{{\mathrm{MF}},\alpha\beta}^{(2),T}(1,2)=\frac{\partial\rho_{\alpha}^{{\mathrm{MF}}}
[\{\bar \nu_{\alpha}-{\rm i}\overline\omega_{\alpha}\}](1)}
{\partial \nu_{\beta}(2)}=\frac{\partial\rho_{\alpha}^{{\mathrm{RS}}}[\{\bar \nu_{\alpha}-{\rm i}\overline\omega_{\alpha}\}](1)}{\partial \nu_{\beta}(2)}.
\label{G-MF}
\end{equation}
$\rho_{\alpha}^{{\mathrm{RS}}}[\{\nu_{\alpha}\}]$ depends on $\nu_{\alpha}$ directly but also through the mean field $\overline\omega_{\alpha}$. Therefore, one has
\begin{eqnarray}
G_{{\mathrm{MF}},\alpha\beta}^{(2),T}(1,2)&=&\left. \frac{\partial\rho_{\alpha}^{{\mathrm{RS}}}
[\{\bar \nu_{\alpha}-{\rm i}\overline\omega_{\alpha}\}](1)}{\partial \nu_{\beta}(2)}\right \vert_{\overline\omega}+\left. \frac{\partial\rho_{\alpha}^{{\mathrm{RS}}}
[\{\bar \nu_{\alpha}-{\rm i}\overline\omega_{\alpha}\}](1)}{\partial \overline\omega_{\gamma}(3)}\right \vert_{\nu_{\gamma}}
\nonumber \\
&&
\times
\frac{\partial\overline
\omega_{\gamma}(3)}{\partial \nu_{\beta}(2).
}
\label{G_field}
\end{eqnarray}
 Taking into account (\ref{2.15a}) and (\ref{G-MF}) we obtain finally 
\begin{equation}
G_{{\mathrm{MF}},\alpha\beta}^{(2),T}(1,2)=G_{{\mathrm{RS}},\alpha\beta}^{(2),T}(1,2)-
\beta G_{{\mathrm{MF}},\alpha\gamma}^{(2),T}(1,3)w_{\gamma\delta}(3,4)G_{{\mathrm{RS}},\delta\beta}^{T}(4,2).
\label{G_field1}
\end{equation} 
(\ref{G_field1}) can be rewritten in a matricial form as \cite{Cai-JSP}
\begin{equation}
\underline{G}_{{\mathrm{MF}}}^{(2),T}(1,2)=\underline{G}_{{\mathrm{RS}}}^{(2),T}(1,2)-
\underline{G}_{{\mathrm{MF}}}^{(2),T}(1,3)\underline{w}(3,4)\underline{G}_{{\mathrm{RS}}}^{(2),T}(4,2),
\label{G_field2}
\end{equation}
where $\underline{G}_{{\mathrm{MF}}({\mathrm{RS}})}^{(2),T}(i,j)$ denotes the matrix of elements
$G_{{\mathrm{MF}}({\mathrm{RS}}),\alpha\beta}^{(2),T}(i,j)$ and $\underline{w}(i,j)$ that of elements
$\beta w_{\alpha\beta}(i,j)$. The formal solution of eq.~(\ref{G_field2}) is then

\begin{eqnarray}
\underline{G}_{\mathrm{MF}}^{(2),T}(1,2)= \left(\underline{1} +\underline{w}
\star \underline{G}^{(2), T}_{\mathrm{RS}}
\right)^{-1} \star  \underline{G}^{(2), T}_{\mathrm{RS}}\left(1,2 \right)
\label{zozo}
\; ,
\end{eqnarray}
where $\underline{1}=\delta_{\alpha\beta}\delta(1,2)$ is the unit operator and the ``$\star$'' denotes a convolution in space.

\subsection{Beyond the MF approximation}

In order to take into account fluctuations we present CVs $\rho_{\alpha}$
and $\omega_{\alpha}$ in the form:
\[
\rho_{\alpha}(1)=\overline\rho_{\alpha}+\delta\rho_{\alpha}(1), \quad
\omega_{\alpha}(1)=\overline\omega _{\alpha}+\delta\omega_{\alpha}(1),
\]
where the quantities with a bar are given by (\ref{2.15a}).

The function $\ln\Xi_{\mathrm RS}[\{\overline\nu_{\alpha};-{\rm i}\omega_{\alpha}\}]$ in (\ref{actionCV}) can be presented  in the form of the cumulant expansion
\begin{eqnarray}
\ln\Xi_{{\mathrm RS}}[\{\overline\nu_{\alpha}-{\rm
i}\omega_{\alpha}\}]=\sum_{n\geq 1}\frac{(-{\rm
i})^{n}}{n!}\sum_{\alpha_{1},\ldots,\alpha_{n}} \int {\rm
d}1\ldots\int {\rm d}n\,
\nonumber\\
\times
{\mathfrak{M}}_{\alpha_{1}\ldots\alpha_{n}}(1,\ldots,n)
\delta\omega_{\alpha_{1}}(1)\ldots\delta\omega_{\alpha_{n}}(n),
\label{2.16}
\end{eqnarray}

where ${\mathfrak{M}}_{\alpha_{1}\ldots\alpha_{n}}(1,\ldots,n)$ is the $n$th cumulant defined by
\begin{equation}
{\mathfrak{M}}_{\alpha_{1}\ldots\alpha_{n}}(1,\ldots,n)=\left. \frac{1}{(-i)^{n}}\frac{\partial^{n}\ln
\Xi_{{\mathrm RS}}[\{\overline\nu_{\alpha}-{\rm i}\omega_{\alpha}\}]}{
\partial\delta\omega_{\alpha_{1}}(1)\ldots\partial\delta\omega_{\alpha_{n}}(n)}\right \vert_{\delta\omega_{\alpha_{i}}=0}.
\label{cumulant_def}
\end{equation}

As is seen from (\ref{cumulant_def}) and (\ref{defcorre}), the $n$th cumulant is equal to the $n$-particle partial truncated (connected) correlation function at $\omega_{\alpha}=\overline\omega_{\alpha}$.
The expressions for the several cumulants given in the Cartesian coordinate phase space are as follows
\begin{equation}
{\mathfrak{M}}_{\alpha_{1}}(1)=\rho_{\alpha_{1}}(1), \label{m_1}
\end{equation}
\begin{equation}
{\mathfrak{M}}_{\alpha_{1}\alpha_{2}}(1,2)=\rho_{\alpha_{1}}(1)\rho_{\alpha_{2}}(2)h_{\alpha_{1}\alpha_{2}}(1,2)+\rho_{\alpha_{1}}(1)\delta_{\alpha_{1}\alpha_{2}}\delta(1,2)],
\label{m_2}
\end{equation}
\begin{eqnarray}
{\mathfrak{M}}_{\alpha_{1}\alpha_{2}\alpha_{3}}(1,2,3)&=&\rho_{\alpha_{1}}(1)\rho_{\alpha_{2}}(2)\rho_{\alpha_{3}}(3)h_{\alpha_{1}\alpha_{2}\alpha_{3}}(1,2,3)
\nonumber \\
&&
+\rho_{\alpha_{1}}(1)\rho_{\alpha_{2}}(2)h_{\alpha_{1}\alpha_{2}}(1,2)
\delta_{\alpha_{1}\alpha_{3}}\delta(1,3)
\nonumber \\
&&
+\rho_{\alpha_{1}}(1)\rho_{\alpha_{3}}(3)h_{\alpha_{1}\alpha_{3}}(1,3)\delta_{\alpha_{1}\alpha_{2}}\delta(1,2)
\nonumber \\
&&
+\rho_{\alpha_{2}}(2)\rho_{\alpha_{3}}(3)h_{\alpha_{2}\alpha_{3}}(2,3)\delta_{\alpha_{1}\alpha_{2}}\delta(1,2)
\nonumber \\
&& +\rho_{\alpha_{1}}(1)\delta_{\alpha_{1}\alpha_{2}}
\delta_{\alpha_{1}\alpha_{3}}\delta(1,2)\delta(1,3), \label{m_3}
\end{eqnarray}
\begin{eqnarray}
{\mathfrak{M}}_{\alpha_{1}\alpha_{2}\alpha_{3}\alpha_{4}}(1,2,3,4)&=&\rho_{\alpha_{1}}(1)\rho_{\alpha_{2}}(2)\rho_{\alpha_{3}}(3)\rho_{\alpha_{4}}(4)h_{\alpha_{1}\alpha_{2}\alpha_{3}\alpha_{4}}(1,2,3,4)
\nonumber \\
&&
+\sum_{i,j,k,l}\rho_{\alpha_{i}}(i)\rho_{\alpha_{j}}(j)\rho_{\alpha_{k}}(k)h_{\alpha_{i}\alpha_{j}\alpha_{k}}(i,j,k)
\nonumber \\
&&
\times\delta_{\alpha_{i}\alpha_{l}}\delta(i,l)+\sum_{i,j,k,l}\rho_{\alpha_{i}}(i)\rho_{\alpha_{k}}(k)h_{\alpha_{i}\alpha_{k}}(i,k)
\nonumber \\
&&
\times\delta_{\alpha_{i}\alpha_{j}}\delta_{\alpha_{k}\alpha_{l}}\delta(i,j)\delta(k,l)+\sum_{i,j,k,l}\rho_{\alpha_{i}}(i)\rho_{\alpha_{l}}(l)
\nonumber \\
&& 
\times h_{\alpha_{i}\alpha_{l}}(i,l)\delta_{\alpha_{i}\alpha_{j}}\delta_{\alpha_{i}\alpha_{k}}\delta(i,j)\delta(i,k)+\rho_{\alpha_{1}}(1)
\nonumber \\
&& 
\times\delta_{\alpha_{1}\alpha_{2}}
\delta_{\alpha_{1}\alpha_{3}}\delta_{\alpha_{1}\alpha_{4}}\delta(1,2)\delta(1,3)\delta(1,4).
\label{m_4}
\end{eqnarray}

In the above formulas  $\rho_{\alpha_{i}}(i)$ is the local density of the $\alpha_{i}$th species in the RS and
$h_{\alpha_{1}\ldots\alpha_{n}}(1,\ldots,n)$ is the $n$-particle partial correlation function of a  $m$-component RS, defined in the GC ensemble
(see eqs. (\ref{3.5c})-(\ref{3.5b})):

\begin{eqnarray}
h_{\alpha_{1}\alpha_{2}}(1,2)&=&g_{\alpha_{1}\alpha_{2}}(1,2)-g_{\alpha_{1}}(1)g_{\alpha_{2}}(2),
\nonumber \\
h_{\alpha_{1}\alpha_{2}\alpha_{3}}(1,2,3)&=&g_{\alpha_{1}\alpha_{2}\alpha_{3}}(1,2,3)-g_{\alpha_{1}\alpha_{2}}(1,2)g_{\alpha_{3}}(3)
\nonumber \\
&&
-g_{\alpha_{1}\alpha_{3}}(1,3) g_{\alpha_{2}}(2)-g_{\alpha_{2}\alpha_{3}}(2,3)g_{\alpha_{1}}(1)
\nonumber \\
&&
+2g_{\alpha_{1}}(1)g_{\alpha_{2}}(2)g_{\alpha_{3}}(3)
\nonumber \\
h_{\alpha_{1}\alpha_{2}\alpha_{3}\alpha_{4}}(1,2,3,4)&=&g_{\alpha_{1}\alpha_{2}\alpha_{3}\alpha_{4}}(1,2,3,4)-\ldots .
\label{h_n}
\end{eqnarray}
In the case of a homogeneous system a Fourier image of the $n$th cumulant can be presented in the form
\begin{eqnarray}
\label{str_fact} {\mathfrak{M}}_{\alpha_{1}\ldots\alpha_{n}}({\bf
k}_{1},\ldots,{\bf k}_{n})&=&(\langle
N_{\alpha_{1}}\rangle,\ldots\langle
N_{\alpha_{n}}\rangle)^{1/n}S_{\alpha_{1}\ldots\alpha_{n}}(k_{1},\ldots,k_{n})
\nonumber \\
&&
\times\delta_{{\bf k}_{1}+\ldots+{\bf k}_{n}},
\end{eqnarray}
where $S_{\alpha_{1}\ldots\alpha_{n}}({\bf k}_{1},\ldots,{\bf
k}_{n})$ is the $n$-particle partial structure factor of the RS.

Substituting (\ref{2.16}) in (\ref{VSS}) one can obtain
\begin{eqnarray}
\Xi\left[\{\nu_{\alpha}\}\right]&=&\Xi_{{\mathrm RS}}\left[\{\overline\nu_{\alpha}-{\rm i}\overline\omega_{\alpha}\}\right]\int \mathcal{D} \delta\rho
\mathcal{D} \delta\omega\;\exp\left\lbrace -\frac{\beta}{2} \left\langle \delta\rho_{\alpha}
\vert w_{\alpha\beta}\vert \delta\rho_{\beta} \right\rangle  
\right.
\nonumber \\
&&\left. +{\rm i}
\left\langle \delta\omega_{\alpha} \vert \delta\rho_{\alpha}\right\rangle
+\sum_{n\geq 2}\frac{(-{\rm
i})^{n}}{n!}\sum_{\alpha_{1},\ldots,\alpha_{n}} \int {\rm
d}1\ldots\int {\rm d}n\,\right.
\nonumber\\
&&\left. \times
{\mathfrak{M}}_{\alpha_{1}\ldots\alpha_{n}}(1,\ldots,n)
\delta\omega_{\alpha_{1}}(1)\ldots\delta\omega_{\alpha_{n}}(n)
\right\rbrace  \; .
\label{Xi_full}
\end{eqnarray} 

Integrating in (\ref{Xi_full}) over $\delta\omega_{\alpha_{i}}(i)$ we have in the homogeneous case
\begin{eqnarray}
\Xi[\{\nu_{\alpha}\}]&=&\Xi_{{\mathrm MF}} \Xi' \int(\mathrm{d}\delta\rho)
\exp\Big\{-\frac{1}{2!}\sum_{\alpha,\beta}\sum_{\bf
k}L_{\alpha\beta}(k)\delta\rho_{{\bf k},\alpha}\delta\rho_{-{\bf
k},\beta} \nonumber \\
&&
+\sum_{n\geq 3}\mathcal{H}_{n}(\delta\rho_{\alpha})\Big\}.
\label{dA.14}
\end{eqnarray}

\paragraph{Gaussian approximation.} In the Gaussian approximation, which corresponds to taking into account in (\ref{dA.14}) only the terms with $n\leq 2$ ($\mathcal{H}_{n}\equiv 0$), we have $L_{\alpha\beta}(k)=C_{\alpha\beta}(k)$, where $C_{\alpha\beta}(k)$
are the Fourier transforms of the  partial direct correlation functions.
After  integrating in (\ref{dA.14})  we arrive at the GPF of a $m$-component system in the random phase approximation (RPA).

Using the Gaussian averages one can develop a loop expansion of $\Xi[\{\nu_{\alpha}\}]$ in the CV representation as it was done recently  for a one-component fluid \cite{caillol_patsahan_mryglod}.

\section{Conclusion}
Using the CV method we have reconsidered the basic relations of
statistical field theory of a multicomponent non-homogeneous fluids that follow from this approach. In contrary to the KSSHE theory \cite{Cai-JSP} the
corresponding CV action depends on two sets scalar fields - field
$\rho_{\alpha}$ connected to the number density of the $\alpha$th species particles and field
$\omega_{\alpha}$ conjugate to $\rho_{\alpha}$. We derive the explicit expressions for the CV field correlations and obtain their relation to the density correlation functions of a multicomponent system.

Contrary to the theories based on the Stratonovich-Hubbard transformation \cite{stratonovich,hubbard}, the CV representation has some important advantages which could
be very useful for more complicate models of fluids. In
particular, it is valid for an arbitrary pair potential (including
a pair interaction $w_{\alpha\beta}(1,2)$ which does not possess an inverse) and
is easily generalized for the case of n-body interparticle
interactions with $n>2$.

\end{document}